\documentclass[preprint,showpacs,preprintnumbers,amsmath,amssymb]{revtex4}

\usepackage{graphicx}% Include figure files
\usepackage{dcolumn}% Align table columns on decimal point
\usepackage{bm}% bold math

\begin{document}
\allowdisplaybreaks{

\preprint{KEK-TH-1313}

\title{Chiral Zero Modes on Intersecting Heterotic 5-branes}% Force line breaks with \\

\author{Tetsuji Kimura}
\email{tetsuji@post.kek.jp}

\author{Shun'ya Mizoguchi}
\altaffiliation[Also at ]{Department of Particle and Nuclear Physics, The Graduate University for Advanced Studies.}%Lines break automatically or can be forced with \\%
\email{mizoguch@post.kek.jp}
\affiliation{%
Theory Center, High Energy \\
Accelerator Research Organization (KEK)\\
Tsukuba, Ibaraki 305-0801, Japan 
}
%\author{}

\date{Dec.8, 2009\\ \phantom{} }% It is always \today, today,

\begin{abstract}
We show that there exist two {\bf 27} and one $\overline{\bf 27}$ of $E_6$, 
net one chiral supermultiplet as zero modes localized on the 
intersecting 5-branes in the $E_8 \times E_8$ heterotic string theory.   
A heterotic background is constructed by the standard embedding 
in the smeared solution, and the Dirac equation is solved explicitly 
on this background. 
It provides, after a compactification of some of the transverse dimensions, 
a Randall-Sundrum II like
brane-world set-up in heterotic string theory. 

%As a by-product, 
%we present  a new proof of anomaly cancellation between those from the 
%chiral matter and the anomaly inflow onto the brane
%%in the $E_8\times E_8$ heterotic theory
%without small instanton. 
\end{abstract}

\maketitle

%\section{\label{sec:Introduction}Introduction%\protect\\ The line
%break was forced \lowercase{via} \textbackslash\textbackslash
}
%
%How the Standard Model emerges in string theory is a long-standing question.
%In early days of string theory, %before D-branes were found, 
%the heterotic string theory \cite{heterotic_string} 
%was considered as a promising candidate for the fundamental theory which would 
%provide a basis for model building. Its miraculous anomaly cancellation allows 
%only two choices (that is, $E_8\times E_8$ and $SO(32)$) of a consistent gauge 
%group, and in Calabi-Yau compactifications (including orbifold and other $1/4$ 
%supersymmetric compactifications in a broad sense) there appear variety of 
%four-dimensional supersymmetric standard-model-like theories with chiral generations. 
%The problem is, however, that the number of such possible compactifications 
%seems too large \cite{Susskind} to find natural necessity for our world to be 
%as observed, despite the remarkable uniqueness of the original theory.

%In the late last century, a conceptually different approach was proposed 
%to realize a four-dimensional world by using D-branes %\cite{D-brane} 
%in type II string theories. 
%The key observation is that 
%two intersecting D-branes can support chiral fermions at the intersection \cite{BDL}.
%Since then many intersecting D-brane models have been built and discussed 
%so far. We refer to the articles \cite{IntersectingDbraneModels} for a review of 
%these developments. Also, inspired by the discovery of D-branes, 
%brane-world models have also been extensively studied 
%as a possible solution to the hierarchy problem and in terms of cosmological 
%model building \cite{ADD, RS}.

In this letter, 
%we propose a new brane-world set-up for $E_6$ GUT 
%model building by using intersecting 5-branes in {\em heterotic string theory}. 
%The 5-branes in heterotic string theory are, of course, not D-branes. They are
%NS5-branes, and unlike D-branes, they are not described by open strings. 
%What makes them hard to deal with is that, near the core of the solution, 
%the geometry is not AdS but an infinite throat 
%where the dilaton diverges linearly. Nevertheless, we can identify what low-energy 
%excitations are on the brane by investigating zero modes of the
%supergravity solution \cite{CHS}. 
%%
%It has been known for some time 
%that on a symmetric 5-brane \cite{CHS} there are 30 $D=6$, ${\cal N}=1$ 
%supermultiplet as zero modes in either of $E_8 \times E_8$ or $SO(32)$ 
%heterotic string theory. In fact, as we explain in section II, they can be regarded as 
%certain Nambu-Goldstone modes associated with various spontaneously broken 
%symmetries of the theory. Therefore, we may expect that, as pions are effectively 
%described %, at low energies, 
%by the chiral model without detailed knowledge of
%QCD, the zero modes on the heterotic 5-branes may also provide enough
%information for low-energy model building, even though their microscopic 
%description (such as little string theory) is not fully understood. 
%The existence of chiral zero modes is also consistent 
%with the anomaly cancellation against an anomaly inflow from the bulk.
%
%
%In order to examine the zero modes on the intersecting system, 
we %first 
construct an intersecting 5-brane solution in the $E_8 \times E_8$ 
heterotic string theory 
by the so-called standard embedding in the known smeared intersecting 
NS5-brane solution of type II theories. We then study the zero modes of the 
relevant Dirac operator on this background and show that there exist three 
localized chiral zero modes, two of which are in the {\bf 27} representation of $E_6$, 
and one in the $\overline{\bf 27}$ representation. They give rise to net {\em one} 
{\bf 27} of massless chiral fermions in the four-dimensional spacetime. 
Therefore, %still being a toy model, 
this is the first example of a brane set-up in heterotic string theory  
that supports  %(after compactifying relatively transverse dimensions)
four-dimensional chiral matter fermions transforming as an $E_6$ gauge multiplet 
\footnote{This corrects the statement made in the earlier version of this paper, 
in which it was erroneously conjectured that the three supermultiplets 
would be of  the same chirality.}.

This letter is a concise version of the paper \cite{KM}, to which the 
reader is referred for further detail.

We start with the neutral smeared solution
\cite{intersecting_solutions}:
\begin{eqnarray}
ds^2&=&\sum_{i=0,7,8,9}\eta_{ij}dx^i dx^j
+h(x^1)^2\sum_{\mu=1,2}\delta_{\mu\nu}dx^\mu dx^\nu
+h(x^1)\sum_{\mu=3,4,5,6}\delta_{\mu\nu}dx^\mu dx^\nu,
\nonumber\\
e^{2\phi}&=&h(x^1)^2,\nonumber\\
H_{\mu\nu\lambda}&=&\left\{
\begin{array}{cl}
\frac{h'}2 ~(=\frac{\xi|x^1|'}2)&\mbox{if $(\mu,\nu,\lambda)=(2,3,4)$,$(2,5,6)$ or their even permutation},\\
-\frac{h'}2~(=-\frac{\xi|x^1|'}2)&\mbox{if $(\mu,\nu,\lambda)=(2,4,3)$,$(2,6,5)$ or their even permutation},\\
0&\mbox{otherwise,}
\end{array}
\right.
\label{intersecting_neutral}
\end{eqnarray}
where
\begin{equation}
h(x^1)=h_0+\xi |x^1|.
\end{equation}
All other components of $H_{MNL}$ vanish. $h_0$ and $\xi$ are real constants. 
The prime ${}'$ denotes 
the differentiation with respect to $x^1$, and $|x^1|'$ is therefore a step function.
This is a solution to equations of motion of the leading-order NSNS-sector 
Lagrangian in type II theories.
%\begin{eqnarray}
%{\cal L}_{NS}&=&
%\frac1{2\kappa^2} \int d^{10}x  \sqrt{-g}e^{-2\phi}
%\left(
%R(\omega)-\frac13 H_{MNP}H^{MNP}+4(\partial_M\phi)^2
%\right).
%\label{LagNSNS}
%\end{eqnarray}
The solution describes a pair of intersecting NS5-branes \cite{SJR,Strominger,CHS} 
stretching in dimensions as shown in Table \ref{tab:table1}. 
These branes are delocalized in the $x^2,x^3,x^4,x^5$ and $x^6$ directions.
Consequently, the solution depends only on $x^1$, and
hence the name ``smeared solution".

\begin{table}[h]
\caption{\label{tab:table1}
Dimensions in which the 5-branes stretch. 
%This is a narrow table which fits into a
%narrow column when using \texttt{twocolumn} formatting. Note that
%REV\TeX~4 adjusts the intercolumn spacing so that the table fills the
%entire width of the column. Table captions are numbered
%automatically. This table illustrates left-aligned, centered, and
%right-aligned columns.  
}
\begin{ruledtabular}
\begin{tabular}{lcccccccccc}
%\footnote{Note a.}
&0%\footnote{Note b.}
&1
&2&3&4&5&6&7&8&9\\
\hline
5-brane1 & $\bigcirc$ & &&
&&$\bigcirc$&$\bigcirc$&$\bigcirc$&$\bigcirc$&$\bigcirc$\\
5-brane2 & $\bigcirc$ &&
&$\bigcirc$&$\bigcirc$&
&&$\bigcirc$&$\bigcirc$&$\bigcirc$\\
\end{tabular}
\end{ruledtabular}
\end{table}

%\subsection{Brane tension and  the harmonic function}
The coefficient $\xi$ in the definition $h(x^1)$ 
is related to the tension of the brane. 
It turns out  that the brane tension $V$  is related to $\xi$ as \cite{KM}
\begin{eqnarray}
\xi&=&-\kappa^2 V h_0^{\frac52},
\label{xi}
\end{eqnarray}
where $\kappa$ is the Newton constant.
%
%Since $e^\phi=h(x^1)$, the sign of $\xi$ strongly affects the dilaton 
%profile. (If $\xi=0$, the solution is reduced to a flat Minkowski space.) 
%We consider the following two cases separately:
%%

\begin{figure}%[h]
{\renewcommand{\arraystretch}{.5}
\begin{tabular}{c@{\hspace{10mm}}c}
\includegraphics[height=0.2\textheight]{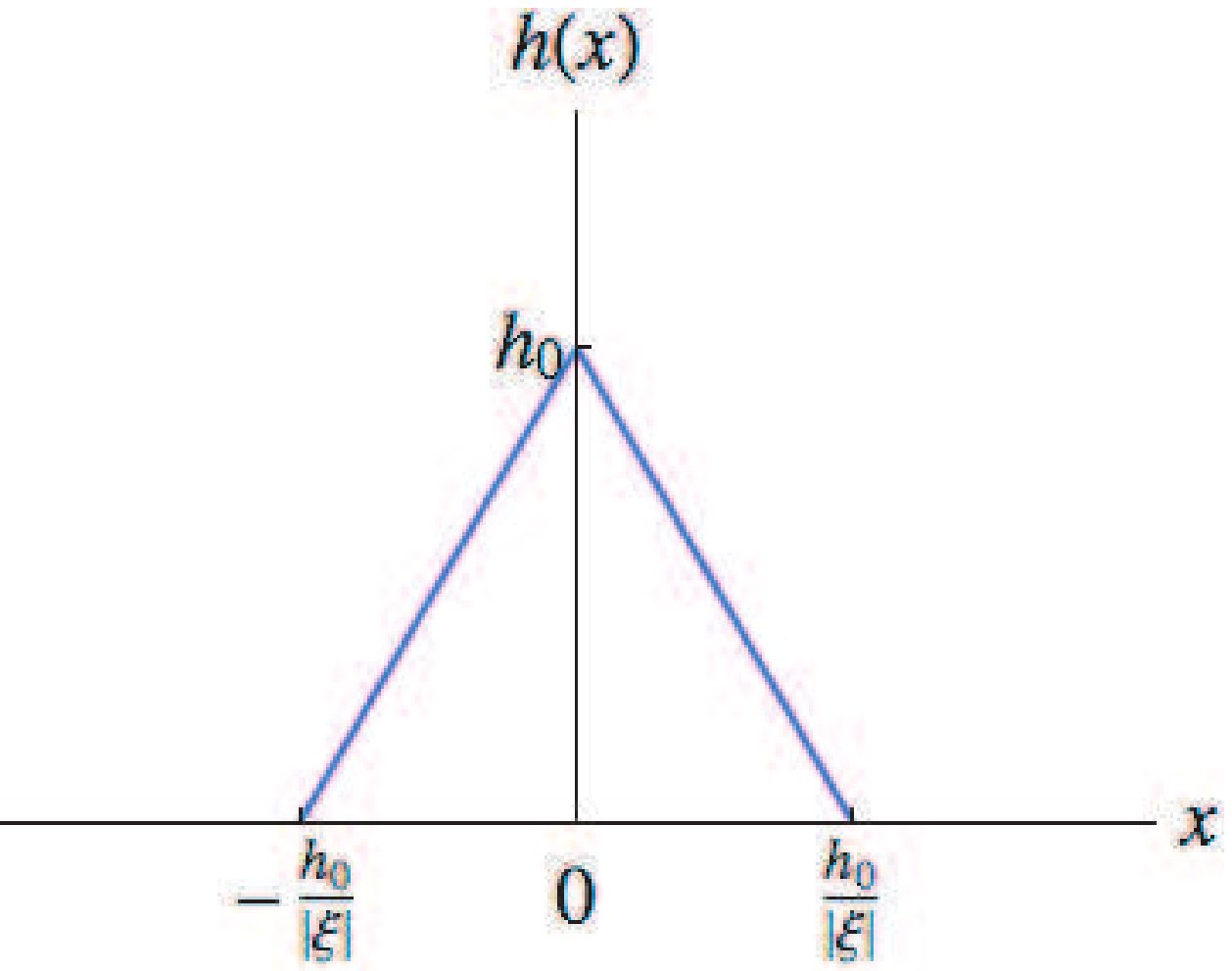}
&
\raisebox{1.5ex}{
\includegraphics[height=0.19\textheight]{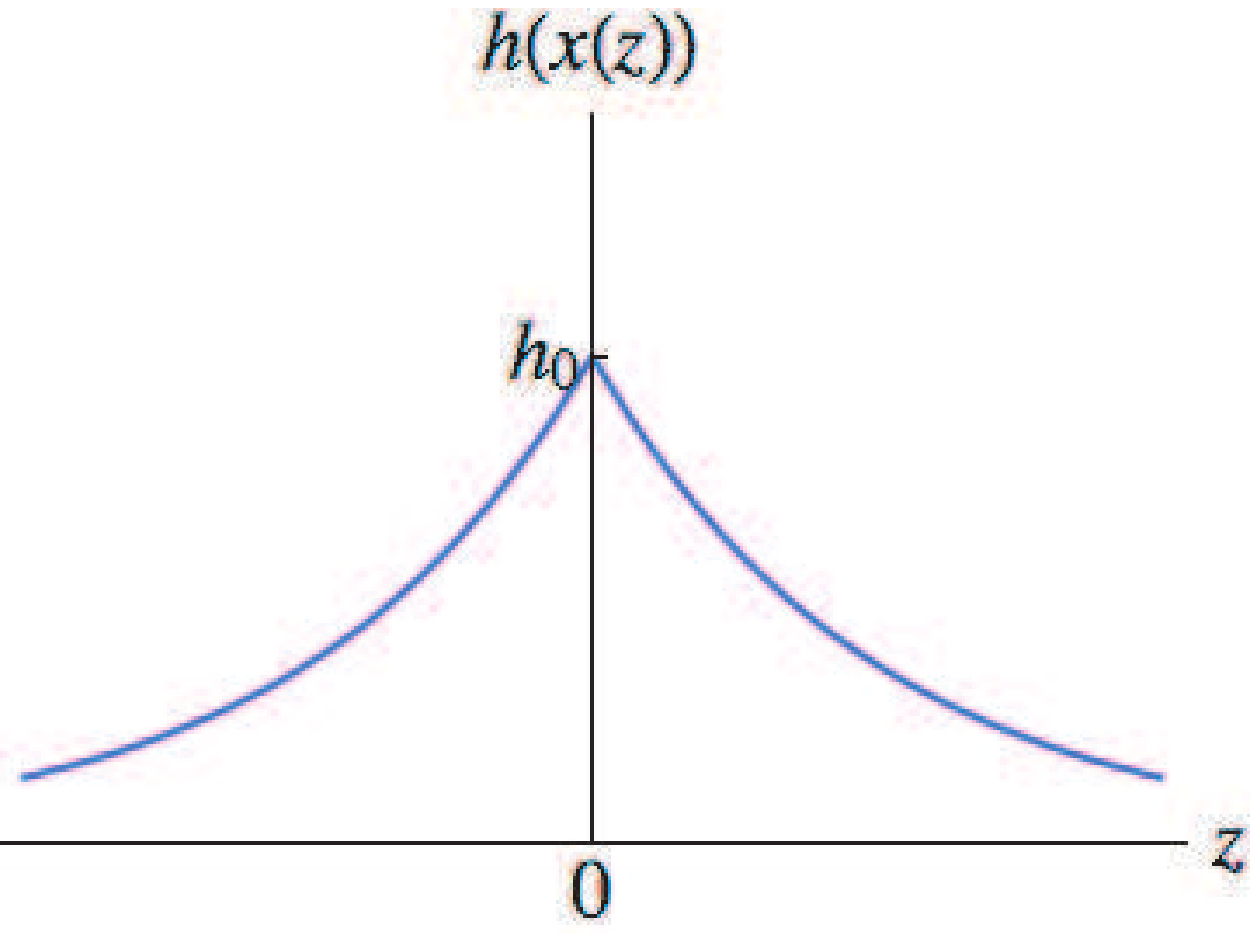}}
\\
(a) & (b)
\end{tabular}
}
\caption{
\label{negative_xi}\sl
$h(x)$ with $\xi<0$.
(a)
The brane has a positive tension. The string coupling  
decreases linearly from a positive value $h_0$, to necessarily cross 
the $x^1$ axis, where the string coupling becomes zero.  We 
identify this point as the ``end of the world''.
(b) 
By a change of the coordinate the points 
$x=\pm\frac{h_0}{|\xi|}$ are mapped to $z=\pm\infty$. The profile of 
$h(x(z))$ becomes similar to the warp factor of the RS II model.}
\end{figure}

%\begin{figure}
%%\includegraphics[width=.26\textwidth]{fig1}% Here is how to import EPS art
%\includegraphics[height=0.2\textheight]{positive_xi.eps}%fig1}% Here is how to import EPS art
%\caption{\label{positive_xi_a} 
%$h(x)$ with $\xi>0$. 
%The brane tension is negative.  Also, 
%the string coupling becomes stronger as one goes away 
%from the branes.}
%
%\includegraphics[height=0.2\textheight]{negative_xi_a.eps}
%\raisebox{1.5ex}{
%\includegraphics[height=0.19\textheight]{negative_xi_b.eps}
%}
%\caption{
%\label{negative_xi} 
%$h(x)$ with $\xi<0$.
%(a)(left):  
%The brane has a positive tension. The string coupling  
%decreases linearly from a positive value $h_0$, to necessarily cross 
%the $x^1$ axis, where the string coupling becomes zero.  We 
%identify this point as the ``end of the world".
%(b)(right): By a change of the coordinate the points 
%$x=\pm\frac{h_0}{|\xi|}$ are mapped to $z=\pm\infty$. The profile of 
%$h(x(z))$ becomes similar to the warp factor of the RS II model.}
%\end{figure}

If $\xi=0$, the solution is reduced to a flat Minkowski space. 
If $\xi >0$,  the brane tension is negative. It is doubtful whether such 
an object may consistently exist in heterotic string theory. 
It is also puzzling that the string coupling becomes stronger as one goes away 
from the branes. Thus, we assume that 
$\xi<0$, then the brane has a positive tension. The string coupling $h(x^1)$ 
is now convex upwards in $x^1$, and 
decreases linearly from a positive value $h_0$, to necessarily cross 
the $x^1$ axis, where the string coupling becomes zero. Beyond that 
point, $h(x^1)$ becomes negative (FIG.1(a)). We 
identify this point as the ``end of the world"; one can 
send this point infinitely far away by the coordinate transformation
\begin{eqnarray}
z&=&-\mbox{sign($x^1$)}\log \frac{h(x^1)}{h_0},
\label{z}
\end{eqnarray}  
where $z$ is the new coordinate (FIG.1(b)). Then the function $h(x^1)$ 
is expressed simply as
\begin{eqnarray}
h&=&h_0~e^{-|z|}.
\end{eqnarray}

Apparently, this looks similar to the Randall-Sundrum (RS) II model \cite{RS}, but 
there are the following differences:
The first is that we have no bulk cosmological constant. 
Secondly, as we see in a moment, there exist chiral zero modes on the branes,
which are in the ${\bf 27}$ representation of $E_6$. This is not an assumption 
but a logical consequence of string theory. 
The final difference is in the warp factor. Unlike the RS models, 
our four-dimensional metric is not warped in the 
string frame \footnote{More curiously, 
although the branes have a positive tension 
as we have derived (\ref{xi}), 
the 4D metric is inversely warped 
(like near the negative tension brane in the RS I model)
in the Einstein frame.}.
%It would be interesting to examine whether gravity or gauge field is localized, 
%but in this paper we will focus only on the localization of chiral fermions.

%\subsection{Intersecting 5-branes in heterotic string theory}
We now construct an intersecting solution in the $E_8 \times E_8$ 
heterotic string theory by the standard embedding, similarly to 
the symmetric 5-brane \cite{CHS}.
The (generalized) spin connections \cite{BdR,KY,TK0704}
of the neutral intersecting background are 
%
%computed as
%%
%\begin{eqnarray}
%(\omega \pm H)_{\mu=1~~\beta}^{~~~~\alpha}
%&=&0,
%\nonumber\\
%(\omega \pm H)_{\mu=2~~\beta}^{~~~~\alpha}
%&=&\frac{h'}h
%\left(
%\begin{array}{cccccc}
%  & -1  &&&&   \\
%1 &   &&&&   \\
%  &&&\pm \frac12   &&\\
%  &&  \mp \frac12 &&&\\
% &&&&&\pm \frac12\\
%  &&&&  \mp \frac12 &
%\end{array}
%\right),
%\nonumber\\
%%
%(\omega \pm H)_{\mu=3~~\beta}^{~~~~\alpha}
%&=&\frac{h'}{2h^{\frac32}}
%\left(
%\begin{array}{cccccc}
%  &   &-1&&&   \\
%  &   &  &\mp 1&&   \\
% ~ 1&&& &&\\
%  & \pm 1&&&&\\
% &&&&~~~&\\
%  &&&&&~~~
%\end{array}
%\right),
%\nonumber\\
%%
%(\omega \pm H)_{\mu=4~~\beta}^{~~~~\alpha}
%&=&\frac{h'}{2h^{\frac32}}
%\left(
%\begin{array}{cccccc}
%  &   &&-1&&   \\
%  &   &\pm1  &&&   \\
% &\mp 1&& &&\\
% ~ 1 &&&&&\\
% &&&&~~~&\\
%  &&&&&~~~
%\end{array}
%\right),
%\nonumber\\
%%
%(\omega \pm H)_{\mu=5~~\beta}^{~~~~\alpha}
%&=&\frac{h'}{2h^{\frac32}}
%\left(
%\begin{array}{cccccc}
%  &   &&&-1&   \\
%  &   &  &&&\mp 1   \\
% &&~~~&&& \\
%  &&& ~~~&&\\
% ~ 1&&&&&\\
%  &\pm 1&&&&
%\end{array}
%\right),
%\nonumber\\
%%
%(\omega \pm H)_{\mu=6~~\beta}^{~~~~\alpha}
%&=&\frac{h'}{2h^{\frac32}}
%\left(
%\begin{array}{cccccc}
%  &   &&&&-1  \\
%  &   &&&\pm1  &   \\
% &&&~~~&& \\
% &&~~~&&&\\
% &\mp 1&&&&\\
%  ~ 1&&&&&
%\end{array}
%\right).
%\label{omega+-H}
%\end{eqnarray}
%%
%The gauge connections are obtained by identifying
identified as the gauge connection %:
%\begin{eqnarray}
$A_\mu^{\alpha\beta}=(\omega+ H)_{\mu}^{~~\alpha\beta}$.
%\end{eqnarray}
The result is
\begin{eqnarray}
A_{\mu=1}^{~~\alpha\beta}&=&0,\nonumber\\
A_{\mu=2}^{~~\alpha\beta}&=&
\frac{h'}{h}
\left(
\begin{array}{ccc}
-s &&\\
&\frac12 s &\\
&& \frac12 s
\end{array}
\right)
=
\frac{h'}{h}
\left(
- \frac{3\lambda_3 + \sqrt{3} \lambda_8}4
\right)
\otimes s ,\nonumber\\
A_{\mu=3}^{~~\alpha\beta}&=&
\frac{h'}{2h^{\frac 32}}
\left(
\begin{array}{ccc}
&-{\bf 1}&\\
{\bf 1}& &\\
&&~~~
\end{array}
\right)
=
\frac{h'}{2h^{\frac 32}}
\left(
-i \lambda_2\right)
\otimes {\bf 1},\nonumber\\
A_{\mu=4}^{~~\alpha\beta}&=&
\frac{h'}{2h^{\frac 32}}
\left(
\begin{array}{ccc}
&-s&\\
-s& &\\
&&~~~
\end{array}
\right)
=
\frac{h'}{2h^{\frac 32}}
\left(
- \lambda_1\right)
\otimes s,\nonumber\\
A_{\mu=5}^{~~\alpha\beta}&=&
\frac{h'}{2h^{\frac 32}}
\left(
\begin{array}{ccc}
&&-{\bf 1}\\
&~~~&\\
~{\bf 1}&&~~~
\end{array}
\right)
=
\frac{h'}{2h^{\frac 32}}
\left(
-i \lambda_5\right)
\otimes {\bf 1},\nonumber\\
A_{\mu=6}^{~~\alpha\beta}&=&
\frac{h'}{2h^{\frac 32}}
\left(
\begin{array}{ccc}
&&-s\\
&~~&\\
-s&&
\end{array}
\right)
=
\frac{h'}{2h^{\frac 32}}
\left(
- \lambda_4\right)
\otimes s,
\label{A_SU(3)}
\end{eqnarray}
where
$\lambda_i$'s ($i=1,\ldots,8$) are the Gell-Mann matrices and   
${\bf 1} \equiv\left(
\begin{array}{cc}
1&\\~~&~~1
\end{array}
\right)$, 
$s \equiv i\sigma_2=\left(
\begin{array}{cc}
&~1\\-1&
\end{array}
\right)$.

The explicit expressions of $\omega_\pm$ %(\ref{omega+-H}) 
show \cite{KM} that both 
are $SU(3)$ connections. %As we did in section II for the symmetric 5-brane, 
Since we have embedded $\omega_+$ into the gauge connection $A$, 
the Bianchi identity
is reduced to $dH=0$, and the solution (\ref{intersecting_neutral}) 
is consistent with it. An $SU(3)$ piece of the $E_8(\times E_8)$ 
gauge connection is given a nonzero expectation value. On the other hand, the fact that
$\omega_-\in SU(3)$ implies that the Killing spinor equations for the gravitino %variation
%(\ref{SUSYgravitino}) 
as well as %, as explained before, 
%the 
gaugino variations 
%(\ref{SUSYgravitino}) 
have a common single Killing spinor. 
It can be checked that this also satisfies the equation for the dilatino SUSY variation 
to lowest order.
%
%\begin{eqnarray}
%\delta\lambda&=&\left(
%-\frac14 \Gamma^M \partial_M \phi +\frac1{24}\Gamma^{MNP}H_{MNP}
%\right)\epsilon~=0.
%\end{eqnarray}
Thus, the background (\ref{intersecting_neutral}) together with (\ref{A_SU(3)}) preserve
1/4 of supersymmetries. 
%It also satisfies the equations of motion (\ref{F_eom}) as it should.  

\begin{figure}
%\includegraphics[width=.26\textwidth]{fig1}% Here is how to import EPS art
%(a)\includegraphics[height=0.24\textheight]{FigE7.eps}
%fig1}% Here is how to import EPS art
%
\includegraphics[height=0.24\textheight]{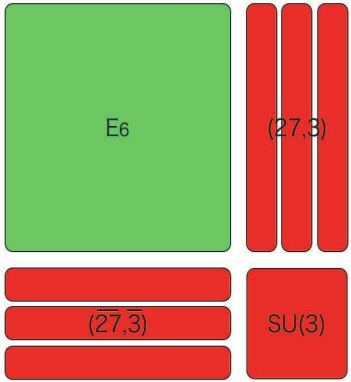}
\caption{\label{FigE7E6} Broken generators which give rise to zero modes.
%(a) The single 5-brane case. (b) The intersecting case.  
}
\end{figure}

%\subsection{Zero modes as Nambu-Goldstone modes on the intersecting 5-branes}
%In the previous subsection we have constructed a smeared solution which 
%describes intersecting 5-branes in the $E_8\times E_8$ heterotic string theory 
%to leading order in $\alpha'$, via the standard embedding, similarly to 
%the way we obtain the symmetric 5-brane. In that case, 
In the case of a single symmetric 5-brane \cite{CHS}, 
the connection $\omega_+$ 
embedded was in $SU(2)$, and the unbroken gauge symmetry was the 
centralizer $E_7$. In the present intersecting case, the connection embedded 
into $E_8$ is in $SU(3)$, and therefore the unbroken gauge symmetry is 
$E_6$.  The adjoint representation of $E_8$ is decomposed into 
\begin{eqnarray}
{\bf 248} &=& ({\bf 78},{\bf 1}) \oplus ({\bf 27},{\bf 3}) 
\oplus (\overline{\bf 27},\overline{\bf 3}) \oplus ({\bf 1},{\bf 8}) 
\label{decomposition}
\end{eqnarray}
as representations of the subalgebra $E_6 \times SU(3)$. 
%
%
%%%%%%%%%%%
%
%Let us examine the moduli of this solution, in particular, the ones 
%coming from the spontaneously broken gauge symmetries. 
%The gauge invariance ensures that any gauge rotation leads to another, 
%possibly different, background which also solves the equations of motion.  
Since the $E_8\times E_8$ gauge field $A_M$ has by construction 
a vev in $SU(3)$, 
%the decomposition 
%%of the adjoint of $E_8$ :
%%\begin{eqnarray}
%$
%{\bf 248}=({\bf 78},{\bf 1}) \oplus
%({\bf 1},{\bf 8}) \oplus
%({\bf 27},{\bf 3}) \oplus
%({\bf \overline{27}},{\bf \overline 3})
%$
%%\end{eqnarray}
%into representations of $E_6 \times SU(3)$
%implies that 
the latter three gauge rotations are the moduli 
(FIG.\ref{FigE7E6}).

Let us focus on the $E_6$ non-singlet moduli. 
%As we saw in the 
%symmetric 5-brane in the previous sections,
The spontaneously broken 
generators in $({\bf 27},{\bf 3}) \oplus
({\bf \overline{27}},{\bf \overline 3})$ give rise to Nambu-Goldstone 
bosons, each of which has one bosonic degree of freedom. 
%The fermionic moduli can be found 
%by supersymmetry. 
On the other hand, since a $D=4$, ${\cal N}=1$ chiral 
supermultiplet needs {\em two} bosonic degrees of 
freedom,
the Nambu-Goldstone bosons which transform as ${\bf 27}$  
and 
${\bf \overline{27}}$ must be combined to form a single ${\cal N}=1$ 
chiral supermultiplet. 
That is, the $E_6$ non-singlet moduli form
{\em  three} chiral supermultiplets in 
the ${\bf 27}$ (or ${\bf \overline{27}}$, but not both) representation of $E_6$.

%%%%%%%%%%%

At first sight, one might think that the argument above would be contradictory  
to the well-known fact in Calabi-Yau compactifications that the number of chiral 
generations are determined by the Dirac index, in which the same decomposition 
(\ref{decomposition}) is used and  {\em one triplet} of zero modes together  
corresponds to {\em one} supermultiplet, and is not counted as three.  
Of course, it is not a contradiction, because what we consider here is not the fermionic zero 
modes of the Dirac operator, but bosonic zero modes of the gauge fields. 
%As we discussed in the previous sections, 
They are not removed by gauge 
transformations, and necessarily exist to cancel the anomaly inflow into each of 
the two intersecting 5-branes \cite{KM}. Each of small gauge rotation generators in 
$({\bf 27},{\bf 3}) \oplus (\overline{\bf 27},\overline{\bf 3}) \oplus ({\bf 1},{\bf 8})$ 
is an independent generator and gives rise to an independent zero mode.  
We also recall that exactly the same way of counting was done in the parallel
symmetric 5-brane case, and was indeed consistent with the index analysis 
\cite{Bellisai}.

%However, it is premature to conclude that these three bosonic zero modes 
%in the $({\bf 27},{\bf 3})$ representation imply three generations, because 
%we have not yet examined the chiralities of their superpartners. We will do this 
%in the next section. In fact, we will see that one of the three possesses the 
%opposite chirality to that the other two have, and hence there is net one generation 
%\footnote{This corrects the statement made in an earlier version of \cite{KM}, 
%in which it was erroneously conjectured that the three supermultiplets 
%would be of 
%the same chirality.}.

%\subsection{\label{sec:chiral_zero_modes}Explicit computation of 
%chiral zero modes%\protect\\ The line
%%break was forced \lowercase{via} \textbackslash\textbackslash
%}
%

Next we examine the fermionic zero modes. 
The ten-dimensional heterotic gaugino equations of motion reads
\begin{eqnarray}
\slash\!\!\!\!D (\omega -\frac 13 H,A)\chi -\Gamma^M \chi \partial_M \phi
+\frac 18 \Gamma^M \gamma^{AB}(F_{AB} + \hat{F}_{AB})
(\psi_M +\frac23 \Gamma_M \lambda)=0,
\end{eqnarray}
where
\begin{eqnarray}
D (\omega -\frac 13 H,A)\chi&\equiv&
\left(
\partial_M +\frac 14(\omega_M^{~~AB}-\frac13 H_M^{~~AB})\Gamma_{AB}
+ \mbox{ad}A_M
\right)\chi.
\end{eqnarray}
$\chi$ is in the adjoint {\bf 248} representation of $E_8$, and $\mbox{ad}A_M
\cdot \chi\equiv {[}A_M,\chi{]}$.
If we set $\psi_M=0$, $\lambda=0$ and
%, it is simplified to
%\begin{eqnarray}
%\slash\!\!\!\!D (\omega -\frac 13 H,A)\chi -\Gamma^M \chi \partial_M \phi
%=0.
%\end{eqnarray}
%Further, 
%if we set 
$\tilde\chi \equiv e^{-\phi} \chi$, then the equation is simplified to 
\begin{eqnarray}
\slash\!\!\!\!D (\omega -\frac 13 H,A)\tilde\chi =0.
\end{eqnarray}
%Below we consider this equation and write $\chi$ without tilde for notational 
%simplicity.

Since there are no nontrivial backgrounds for the four-dimensional 
$i=0,7,8,9$ directions, 
\begin{eqnarray}
\Gamma^i \partial_i \tilde{\chi} + \Gamma^\mu D_\mu(\omega -\frac 13 H,A)\tilde{\chi} =0.
\end{eqnarray}
If $\tilde{\chi}=\tilde{\chi}_{4D}\otimes \tilde{\chi}_{6D}$, the second term is 
regarded as the mass term for the four-dimensional spinor $\tilde{\chi}_{4D}$. 
We are interested in the zero modes of this Dirac operator 
$\Gamma^\mu D_\mu(\omega -\frac 13 H,A)$.

We fix the $SO(6)$ gamma matrices as %in the chiral representation are 
\begin{eqnarray}
\gamma_1&=&\sigma_2 \otimes {\bf 1}\otimes {\bf 1},~~
\gamma_2~=~\sigma_1 \otimes \sigma_1 \otimes {\bf 1},\nonumber\\
\gamma_3&=&\sigma_1 \otimes \sigma_2 \otimes {\bf 1},~~
\gamma_4~=~\sigma_1 \otimes \sigma_3 \otimes \sigma_1,\nonumber\\
\gamma_5&=&\sigma_1 \otimes \sigma_3 \otimes \sigma_2,~~
\gamma_6~=~\sigma_1 \otimes \sigma_3 \otimes \sigma_3.
\end{eqnarray}
The six-dimensional chiral operator is
%\begin{eqnarray}
$\gamma_\sharp\equiv-i \gamma_1 \gamma_2 
%\gamma_3\gamma_4 \gamma_5 
\cdots
\gamma_6 
=\sigma_3 \otimes {\bf 1}\otimes {\bf 1}$.
%\end{eqnarray}
%
For $SO(9,1)$ gamma matrices, we take
%\begin{eqnarray}
$\Gamma^a=\gamma_{4D}^a \otimes {\bf 1}_8$
$(a=0,7,8,9)$, %\nonumber\\
$\Gamma^\alpha=\gamma_{4D}^\sharp \otimes \gamma_\alpha$ ~~~
$(\alpha=1,\ldots,6)$,
%\end{eqnarray}
where $\gamma_{4D}^a$'s $(a=0,7,8,9)$ are the ordinary $SO(3,1)$ gamma matrices 
in the chiral representation.
%:
%\begin{eqnarray}
%\gamma_{4D}^0&=&i \sigma_2 \otimes {\bf 1},\nonumber\\
%\gamma_{4D}^7&=&\sigma_1 \otimes \sigma_1,\nonumber\\
%\gamma_{4D}^8&=&\sigma_2 \otimes \sigma_2,\nonumber\\
%\gamma_{4D}^9&=&\sigma_3 \otimes \sigma_3,\nonumber\\
%\gamma_{4D}^\sharp&\equiv&-i \gamma_{4D}^0 \gamma_{4D}^7
%\gamma_{4D}^8 \gamma_{4D}^9\nonumber\\
%&=&\sigma_3 \otimes {\bf 1}.
%\end{eqnarray}
%
The ten-dimensional chirality is 
%\begin{eqnarray}
$\Gamma_{11}\equiv%-\Gamma^0\Gamma^7\Gamma^8\Gamma^9
%\cdot \Gamma^1\cdots\Gamma^6
%\nonumber\\
%&=&
\gamma_{4D}^\sharp \otimes \gamma_\sharp
%\nonumber\\
=(\sigma_3\otimes {\bf 1}) \otimes (\sigma_3\otimes {\bf 1}\otimes {\bf 1})$.
%\end{eqnarray}

Now we consider the Dirac equation
\begin{eqnarray}
\Gamma^\mu D_\mu (\omega-\frac13 H, A)\tilde{\chi} =0.
\end{eqnarray}
%The gaugino $\chi$ is (Majorana-)Weyl, so for definiteness we assume 
%$\Gamma_{11} \chi = + \chi$. 
%
The 16-component $SO(9,1)$ (Majorana-)Weyl spinor $\chi$ (or $\tilde{\chi}$)
is decomposed in terms of $SO(3,1)$ and $SO(6)$ spinors as
%\begin{eqnarray}
${\bf 16} =({\bf 2}_+,{\bf 4}_+) \oplus ({\bf 2}_-,{\bf 4}_-)$,
%\end{eqnarray}
where the subscripts are the $SO(3,1)$ and $SO(6)$ chiralities, 
$\gamma_{4D}^\sharp$ and $\gamma_\sharp$, respectively. 
Since $\tilde{\chi}$ is Majorana (but complex in this representation), 
the $({\bf 2}_+,{\bf 4}_+)$ 
and $({\bf 2}_-,{\bf 4}_-)$ components are not independent but
are transformed each other by a charge conjugation. 
%
%Therefore it is enough to consider the $({\bf 2}_+,{\bf 4}_+)$ component of $\tilde{\chi}$,
%and the $({\bf 2}_-,{\bf 4}_-)$ component of the zero modes can be obtained 
%by charge conjugation.
%

As $\Gamma^\mu D_\mu (\omega-\frac13 H, A)$ is $SO(3,1)$
diagonal, it is enough to consider
\begin{eqnarray}
\gamma^\mu D_\mu (\omega-\frac13 H, A)\tilde{\chi}_{6D}&=&0,
\label{reduced_Dirac_eq}
\end{eqnarray}
with the understanding that each 
component of $\tilde{\chi}_{6D}$ is accompanied by a two-component $SO(3,1)$ Weyl spinor
with a correlated chirality ($\gamma_\sharp \gamma_{4D}^\sharp = +1$).

On the other hand, we are interested in the gaugino zero modes in $({\bf 27}, {\bf 3})$
or $(\overline{\bf 27}, \overline{\bf 3})$ in the decomposition $E_8  \supset E_6 \times SU(3)$
of ${\bf 248}$. The gauge connections $A_M$ take only nonzero values in the 
$SU(3)$ subalgebra, and we look for the zero modes $\tilde{\chi}_{6D}$ transforming as a triplet,
either ${\bf 3}$ or $\overline{\bf 3}$, of $SU(3)$.

Since $\gamma^\alpha$'s are in the form:
\begin{eqnarray}
\gamma^1&=&
\left(
\begin{array}{cc}
&-i {\bf 1}\otimes {\bf 1}\\
-i {\bf 1}\otimes {\bf 1}&
\end{array}
\right),
\nonumber\\
\gamma^{\tilde{\alpha}}&=&
\left(
\begin{array}{cc}
&\tilde{\gamma}^{\tilde{\alpha}}\\
\tilde{\gamma}^{\tilde{\alpha}}&
\end{array}
\right)~~~(\tilde{\alpha}=2,\ldots,6),
\end{eqnarray}
and $\omega_{\mu}^{~\alpha\beta}$, $H_{\mu}^{~\alpha\beta}$
and $A_{\mu}^{~\alpha\beta}$ all vanish if $\mu=1$, 
(\ref{reduced_Dirac_eq}) is reduced to two independent differential 
equations
\begin{eqnarray}
\frac ih\frac d{d x^1} {\tilde{\chi}}^{+}_{6D} + M^+ {\tilde{\chi}}^{+}_{6D} =0, \label{M+equation}\\
\frac ih\frac d{d x^1} {\tilde{\chi}}^{-}_{6D} - M^- {\tilde{\chi}}^{-}_{6D} =0,\label{M-equation}
\end{eqnarray}
where ${\tilde{\chi}}^{\pm}_{6D}$ is the upper and lower components having definite chiralities:
%\begin{eqnarray}
$\tilde{\chi}_{6D}=\left(
\begin{array}{c}
{\tilde{\chi}}^{+}_{6D}\\
{\tilde{\chi}}^{-}_{6D}
\end{array}
\right)$.
%\end{eqnarray}
${\tilde{\chi}}^{+}_{6D}$ (${\tilde{\chi}}^{-}_{6D}$) is a {\bf 4} $SO(6)$ Weyl spinor, and each of the 
four components is a triplet of $SU(3)$.
Thus $M^+$ ($M^-$) is a $(4\times 3=)$ 12-by-12 matrix, given explicitly by

\begin{eqnarray}
\left(
\begin{array}{cc}
&M^-\\ M^+&
\end{array}
\right)&\equiv&\frac{h'}{h^2}
\left(
{\scriptsize
\left(
\begin{array}{llllllll}
0 & 0 & 0 & 0 & -\frac{3 i}{2} & -\frac{i}{4} & 0 & -\frac{i}{4} \\
0 & 0 & 0 & 0 & -\frac{i}{4} & -\frac{3 i}{2} & -\frac{i}{4} & 0 \\
0 & 0 & 0 & 0 & 0 & -\frac{i}{4} & -\frac{3 i}{2} & -\frac{i}{4} \\
0 & 0 & 0 & 0 & -\frac{i}{4} & 0 & -\frac{i}{4} & -\frac{3 i}{2} \\
\frac{3 i}{2} & -\frac{i}{4} & 0 & -\frac{i}{4} & 0 & 0 & 0 & 0 \\
-\frac{i}{4} & \frac{3 i}{2} & -\frac{i}{4} & 0 & 0 & 0 & 0 & 0 \\
0 & -\frac{i}{4} & \frac{3 i}{2} & -\frac{i}{4} & 0 & 0 & 0 & 0 \\
-\frac{i}{4} & 0 & -\frac{i}{4} & \frac{3 i}{2} & 0 & 0 & 0 & 0
\end{array}
\right)} 
\otimes {\bf 1}_3
\right.
\nonumber\\
&&\left.
+
{\scriptsize
\left(
\begin{array}{llllllll}
0 & 0 & 0 & 0 & -\frac{s \lambda _4}{2} & \frac{-s \lambda _1-\lambda _5}2
  & \frac{-2 \lambda _2-s \lambda_9}4 &
  0 \\
0 & 0 & 0 & 0 & \frac{\lambda _5-s \lambda _1}2 & \frac{s \lambda _4}{2} &
  0 & \frac{-2 \lambda _2-s \lambda_9}4
  \\
0 & 0 & 0 & 0 & \frac{2 \lambda _2-s \lambda_9}4 & 0 & \frac{s \lambda _4}{2} & 
\frac{s \lambda _1+\lambda_5}2 \\
0 & 0 & 0 & 0 & 0 & \frac{2 \lambda _2-s \lambda_9}4 & 
\frac{s \lambda _1-\lambda _5}2 &
 -\frac{s \lambda_4}{2} \\
-\frac{s \lambda _4}{2} & \frac{-s \lambda _1-\lambda _5}2 & \frac{
-2 \lambda _2-s \lambda_9}4& 0 & 0 & 0 & 0
  & 0 \\
\frac{\lambda _5-s \lambda _1}2& \frac{s \lambda _4}{2} & 0 & \frac{
-2 \lambda _2-s \lambda_9}4& 0 & 0 & 0 & 0
  \\
\frac{2 \lambda _2-s \lambda_9}4 & 0 &
  \frac{s \lambda _4}{2} & \frac{s \lambda _1+\lambda _5}2 & 0 & 0 & 0 & 0
  \\
0 & \frac{2 \lambda _2-s \lambda_9 }4&
  \frac{s \lambda _1-\lambda _5}2 & -\frac{s \lambda _4}{2} & 0 & 0 & 0 & 0
\end{array}
\right)
}
\right),\nonumber\\
\end{eqnarray}
where 
%$\lambda_i$'s ($i=1,\ldots,8$) are the GellMann matrices and   
$\lambda_9\equiv 3 \lambda _3+\sqrt{3}\lambda_8$. 
In identifying the spin connection as an $SU(3)$ gauge connection, 
$s=\left(
\begin{array}{cc}
&~1\\-1&
\end{array}
\right)$ can either be mapped to $i$, or to $-i$, 
and depending on this choice, the $SU(3)$ gauge connection 
matrix becomes one
in the {\bf 3} representation, or in the $\overline{\bf 3}$ representation.

%As we already mentioned, 
Since ${\tilde{\chi}}^{+}_{6D} $ and ${\tilde{\chi}}^{-}_{6D} $ 
are not independent, we have only to solve the equation (\ref{M+equation}),
and the solutions to (\ref{M-equation}) may then be obtained by a charge 
conjugation. To solve (\ref{M+equation}), we diagonalize $M^+$ to obtain 
its eigenvalues. Let $i\lambda$ be an eigenvalue of the {\em constant} matrix 
$\left(\frac {h'}{h^2}\right)^{-1}M^+$, and $\psi_\lambda(x^1)$ 
be the corresponding eigenfunction, then they satisfy
\begin{eqnarray}
\frac i{h} \psi'_\lambda + i\lambda \frac{h'}{h^2}\psi_\lambda&=&0.
\end{eqnarray}
This is solved to give
\begin{eqnarray}
\psi_\lambda(x^1)&=&\mbox{const.} (h(x^1))^{-\lambda}.
\end{eqnarray}
Thus, for each eigenvalue,  there exists a zero mode of the Dirac operator.
Since $\xi$ is negative for positive tension, if $\lambda<1$, the mode is 
localized near $x^1=0$, while if $\lambda \geq 1$, it is not localized, being 
either non-normalizable or localized rather at ``infinity" $x^1=\pm\frac{h_0}{|\xi|}$.

The list of eigenvalues of $\left(\frac {h'}{h^2}\right)^{-1}M^+$ is as follows:
If $s=+i$, the eigenvalues are
\begin{eqnarray}
\left\{2 i,\frac{3 i}{2},\frac{3 i}{2},i,-i,i,i,i,\frac{3 i}{2},\frac{3 i}{2},\frac{7
  i}{2},\frac{7 i}{2}\right\},
\label{eigenvalues+i}
\end{eqnarray}
while
if $s=-i$, they are
\begin{eqnarray}
\left\{2 i,\frac{3 i}{2},\frac{3 i}{2},i,2 i,4 i,2 i,2 i,-\frac{i}{2},-\frac{i}{2},\frac{3
  i}{2},\frac{3 i}{2}\right\}.
\label{eigenvalues-i}
\end{eqnarray}

We can clearly see an asymmetry between (\ref{eigenvalues+i}) and 
(\ref{eigenvalues-i}), in particular that the former has only one negative 
(times imaginary unit) eigenvalue, while the latter has two negative 
eigenvalues. 
%The (unnormalized) profiles of the eigenfunctions are shown 
%in Figure.
Assuming that the branes have positive tension so that the function $h(x)$ 
has the profile shown in FIG.2, these are the only modes whose profiles 
have a peak at $x^1=0$ or $z=0$ in the coordinate (\ref{z}). 
The same is also true for the original gaugino variable $\chi = h \tilde{\chi}$
(although the modes with $\lambda=+1$ then become constant).
This result implies that there are indeed three localized modes, and two of 
them are in one (say,  ({\bf 27},{\bf 3})) representation, and the rest belongs to 
the other (($\overline{\bf 27},\overline{\bf 3}$)) representation.

\begin{acknowledgments}
We would like to thank 
Tohru Eguchi, Satoshi Iso, Hikaru Kawai, Taichiro Kugo, Hiroshi Kunitomo, 
Nobuyoshi Ohta and Shigeki Sugimoto for illuminating discussions.
We are also grateful to 
Keiichi Akama, 
Masafumi Fukuma,
Machiko Hatsuda,
Takeo Inami, 
Akihiro Ishibashi, 
Katsushi Ito, 
Katsumi Itoh,
Hiroshi Itoyama,
Yoichi Kazama,
Yoshio Kikukawa,
Yoshihisa Kitazawa, 
Hideo Kodama, 
Nobuhiro Maekawa, 
Nobuhito Maru,  
Yoji Michishita, 
Muneto Nitta,
Kazutoshi Ohta, 
Soo-Jong Rey,
Tomohiko Takahashi, 
Shinya Tomizawa, 
Tamiaki Yoneya
and 
Kentaro Yoshida
for discussions and comments.
We thank YITP for hospitality and support 
during the workshop: ``Branes,
Strings and Black Hole'', where part of this work was done. 
This work is supported by 
%the Ministry of Education,
%Culture, Sports, Science and Technology of Japan. 
Grant-in-Aid
for Scientific Research (C) \#20540287-H20 from
The Ministry of Education, Culture, Sports, Science
and Technology of Japan.

\end{acknowledgments}

%\end{document}


\begin{thebibliography}{99}
\bibitem{KM}
T.~Kimura and S.~Mizoguchi,
 ``{\it Chiral Generations on Intersecting 5-branes in Heterotic String Theory},''
 arXiv:0912.1334 [hep-th], KEK-TH-1341.
 %%CITATION = ARXIV:0912.1334;%%


\bibitem{intersecting_solutions}
R.~Argurio, F.~Englert and L.~Houart,
%``Intersection rules for p-branes,''
Phys.\ Lett.\  B {\bf 398}, 61 (1997)
[arXiv:hep-th/9701042]\\
N.~Ohta,
%``Intersection rules for non-extreme p-branes,''
Phys.\ Lett.\  B {\bf 403}, 218 (1997)
[arXiv:hep-th/9702164].

\bibitem{SJR}
S.~J.~Rey,
%``THE CONFINING PHASE OF SUPERSTRINGS AND AXIONIC STRINGS,''
 Phys.\ Rev.\  D {\bf 43}, 526 (1991);
 %%CITATION = PHRVA,D43,526;%%
``{\it Axionic String Instantons And Their Low-Energy Implications,}''
%Soo-Jong Rey, (UC, Santa Barbara) . UCSB-TH-89/49, Nov 1989. 10pp.
Invited talk given at Workshop on Superstrings and Particle Theory, 
Tuscaloosa, Alabama, Nov 8-11, 1989. 
Published in Tuscaloosa Workshop 1989:0291-300;
``{\it On string theory and axionic strings and instantons,}''
%Soo-Jong Rey, (SLAC) . SLAC-PUB-5659, Sep 1991. 6pp. 
Presented at Particle and Fields '91 Conf., Vancouver, Canada, Aug 18-22, 1991. 
Published in DPF Conf.1991:0876-881. 

\bibitem{Strominger}
A.~Strominger,
 %``Heterotic solitons,''
 Nucl.\ Phys.\  B {\bf 343}, 167 (1990)
 [Erratum-ibid.\  B {\bf 353}, 565 (1991)].
 %%CITATION = NUPHA,B343,167;%%


\bibitem{CHS}
C.~G.~Callan, J.~A.~Harvey and A.~Strominger,
%``World sheet approach to heterotic instantons and solitons,''
Nucl.\ Phys.\  B {\bf 359}, 611 (1991);
%%CITATION = NUPHA,B359,611;%%
%\bibitem{CHS367}
%C.~G.~Callan, J.~A.~Harvey and A.~Strominger,
%``Worldbrane actions for string solitons,''
Nucl.\ Phys.\  B {\bf 367}, 60 (1991).
%%CITATION = NUPHA,B367,60;%%
%

\bibitem{RS}
L.~Randall and R.~Sundrum,
 %``A large mass hierarchy from a small extra dimension,''
 Phys.\ Rev.\ Lett.\  {\bf 83}, 3370 (1999)
 [arXiv:hep-ph/9905221];
 %%CITATION = PRLTA,83,3370;%%
%
%\bibitem{RS2}
%L.~Randall and R.~Sundrum,
 %``An alternative to compactification,''
 Phys.\ Rev.\ Lett.\  {\bf 83}, 4690 (1999)
 [arXiv:hep-th/9906064].
 %%CITATION = PRLTA,83,4690;%%

\bibitem{BdR}
E.~A.~Bergshoeff and M.~de Roo,
%``The Quartic Effective Action Of The Heterotic String And Supersymmetry,''
Nucl.\ Phys.\  B {\bf 328}, 439 (1989);
%%CITATION = NUPHA,B328,439;%%


\bibitem{KY}
T.~Kimura and P.~Yi,
%``Comments on heterotic flux compactifications,''
JHEP {\bf 0607}, 030 (2006)
[arXiv:hep-th/0605247].
%%CITATION = JHEPA,0607,030;%%



\bibitem{TK0704}
T.~Kimura,
 %``Index Theorems on Torsional Geometries,''
 JHEP {\bf 0708}, 048 (2007)
 [arXiv:0704.2111 [hep-th]].
 %%CITATION = JHEPA,0708,048;%%





\bibitem{Bellisai}
 D.~Bellisai,
 %``Fermionic zero-modes around string solitons,''
 Nucl.\ Phys.\  B {\bf 467}, 127 (1996)
 [arXiv:hep-th/9511198].
 %%CITATION = NUPHA,B467,127;%%

\end{thebibliography}
\end{document}